\documentclass[fleqn,usenatbib]{mnras}

\usepackage{newtxtext,newtxmath}

\usepackage[T1]{fontenc}

\DeclareRobustCommand{\VAN}[3]{#2}
\let\VANthebibliography\thebibliography
\def\thebibliography{\DeclareRobustCommand{\VAN}[3]{##3}\VANthebibliography}

\usepackage{graphicx}	
\usepackage{subcaption}
\usepackage{amsmath}	
\usepackage{bm}
\usepackage{color}

\newcommand{\mmat}[1]{{\mathbf{#1}}}
\newcommand{\mP}{\mathcal{P}}
\newcommand{\mD}{\mathcal{D}}
\newcommand{\mO}{\mathcal{O}}

\newcommand{\mL}{\mathcal{L}}

\newcommand{\ia}{\text{IA}}
\newcommand{\mvec}[1]{{\bm{#1}}}
\newcommand{\mTheta}{\mvec{\Theta}}

\newcommand{\cocoa}{{\sc cocoa}~}
\newcommand{\cosmolike}{{\sc cosmolike}~}
\newcommand{\cobaya}{{\sc cobaya}~}
\newcommand{\class}{{\sc CLASS}~}

\newcommand{\camb}{{\sc CAMB}~}
\newcommand{\ns}{$n_\text{s}$~}

\title[Iterative GP\&NN emulator]{Accelerating cosmological inference with Gaussian processes and neural networks - an application to LSST Y1 weak lensing and galaxy clustering}

\author[S. S. Boruah, et. al.]{Supranta S. Boruah$^{1}$\thanks{Contact e-mail: \href{mailto:ssarmabo@email.arizona.edu}{ssarmabo@email.arizona.edu}}, 
Tim Eifler$^{1}$,
Vivian Miranda$^{1,2}$,
Sai Krishanth P.M. $^1$
\\
$^{1}$ Department of Astronomy and Steward Observatory, University of Arizona, 933 N Cherry Ave, Tucson, AZ 85719, USA \\
$^{2}$ C. N. Yang Institute for Theoretical Physics, Stony Brook University, Stony Brook, NY 11794 
}

\date{Accepted XXX. Received YYY; in original form ZZZ}

\pubyear{2021}
\begin{document}
\label{firstpage}
\pagerange{\pageref{firstpage}--\pageref{lastpage}}
\maketitle

\begin{abstract}
Studying the impact of systematic effects, optimizing survey strategies, assessing tensions between different probes and exploring synergies of different data sets require a large number of simulated likelihood analyses, each of which cost thousands of CPU hours. In this paper, we present a method to accelerate cosmological inference using emulators based on Gaussian process regression and neural networks. We iteratively acquire training samples in regions of high posterior probability which enables accurate emulation of data vectors even in high dimensional parameter spaces. We showcase the performance of our emulator with a simulated 3x2 point analysis of LSST-Y1 with realistic theoretical and systematics modelling. We show that our emulator leads to high-fidelity posterior contours, with an order of magnitude speed-up. Most importantly, the trained emulator can be re-used for extremely fast impact and optimization studies. We demonstrate this feature by studying baryonic physics effects in LSST-Y1 3x2 point analyses where each one of our MCMC runs takes approximately 5 minutes. This technique enables future cosmological analyses to map out the science return as a function of analysis choices and survey strategy.
\end{abstract}


\section{Introduction}

Cosmology is entering an exciting new phase with the advent of the Stage-IV surveys such as the Vera C. Rubin Observatory's Legacy Survey of Space and Time \citep[LSST\footnote{\url{https://www.lsst.org/}},][]{LSST19}, Euclid\footnote{\url{https://sci.esa.int/web/euclid}} \citep{laa11}, the Spectro-Photometer for the History of the Universe, Epoch of Reionization, and Ices Explorer \citep[SPHEREx\footnote{\url{http://spherex.caltech.edu/}},][]{dba14}, the Nancy G. Roman Space Telescope \citep[NGRST\footnote{\url{https://roman.gsfc.nasa.gov/}},][]{sgb15}, and the ground based \citep[DESI \footnote{\url{https://www.desi.lbl.gov/}}, ][]{DESI16} and Prime Focus Spectrograph \citep[PFS\footnote{\url{https://pfs.ipmu.jp/}}, ][]{tec14} surveys. These surveys will probe the Universe at an exquisite precision and can potentially revolutionize our understanding of fundamental physics.  

Most current cosmological analyses rely on comparing an observed summary statistic (e.g, 2 point functions) to a theoretical model to constrain the parameters of interest. The summary statistic is binned, usually as a function of projected distance and redshift of the objects, and concatenated into a vector. While the data vector is computed once from the observed catalogs, the calculation of the model vector enters every step of a Markov Chain Monte Carlo (MCMC) analysis. The model vector evaluation involves computationally expensive steps such as the execution of Boltzmann codes such as \class \citep{Blas2011} or \camb \citep{Lewis2002, Howlett2012} and multidimensional numerical integration. Depending on the complexity of the physics and systematics model and the number of cosmological probes considered, the computing time per model vector is of order 10 seconds. 

For each MCMC run, $\mO(10^5$-$10^6$) evaluations required, leading to $\mO(10^4)$ CPUh per likelihood analysis. Given that the actual analysis of the data vector will only be conducted a few times, e.g. for predefined science cases and analysis choices, these final likelihood analyses do not pose a severe computational problem. However, the earlier phase of simulated likelihood analyses that only involve synthetic data have to be conducted much more frequently and the associated computational cost motivate the content of this paper. Across testing and verification of the main code routines, determining scale cuts, impact studies of systematics, understanding of parameter priors, a large number of simulated likelihood analyses are required \citep{Hikage2019, Krause2021}. These need to be repeated for the different science cases and for the different probe combinations. In addition, simulated analyses are needed to optimize the survey strategy of individual surveys \citep[e.g,][]{SimonsObservatory, Eifler2021a} and when considering survey combinations \citep[e.g.,][]{LSSTEuclidWFIRST, Eifler2021b}. Similarly, assessing tensions when comparing cosmological probes \citep[e.g,][]{Raveri2019, Miranda:2020lpk} requires hundreds of MCMC runs to properly quantify whether the observed tension level indicates differences in the underlying physics model or whether it can be explained by regular noise in the data. 

As the complexity of modelling grows for future surveys, the search space for optimization will increase massively. The overall number of simulated likelihood analyses that are necessary to optimize the science return of future cosmological experiments can easily reach several thousands. Therefore, accelerating these analyses is a critical aspect for Stage IV surveys.

A popular method of speeding up the theoretical modelling of cosmological observables is using emulators. We can categorize the various emulators in the literature according to the method used. Polynomial-based emulators were among the first to be used for cosmological analyses. Examples of such emulators include the {\sc pico} \citep{Fendt2007} and {\sc cmbwarp} \citep{Jimenez2004} softwares that were used for Cosmic Microwave Background (CMB) data analysis. More recently, Polynomial Chaos Expansion (PCE) has been used to emulate the nonlinear matter power spectrum \citep{Knabenhans2019} and clustering statistics from hybrid $N$-body-perturbation theory model \citep{Kokron2021}. 

Perhaps the most widely used method of emulation in cosmological applications is Gaussian Process (GP) regression. GP emulators have been used for various cosmological applications such as emulating the matter power spectrum \citep{Heitmann2009, Mootoovaloo2022, Ho2022}, halo mass function \citep{McClintock2019}, Lyman-$\alpha$ forest flux power spectrum \citep{Rogers2019, Pedersen2021} and non-Gaussian weak lensing statistics \citep{Liu2015, Liu2019, Marques2019}. 

Recently, neural networks (NN) have also become a popular method of emulation. {\sc cosmonet} \citep{Auld2007, Auld2008} used NNs for parameter estimation from CMB data. Other notable uses of neural network emulators in cosmology and astrophysics include \citet{Alsing2020, ManriqueYus2020, Arico2021, Kobayashi2021, SpurioMancini2022, DonaldMcCann2022}.  

Different emulation techniques have different strengths and weaknesses. For example, the polynomial-based emulators work very well in low dimensions and for smooth functions. But these emulators do not scale well with dimensions. On the other hand Gaussian process and neural networks both work well in high dimensions. While neural networks perform better with a large training data set, adding more data slows down GP regression. Conversely, Gaussian processes can work well with a limited training data set and is more robust to the details of training compared to neural networks. 

The above mentioned emulators are global emulators in the sense that the training sample is drawn from a wide region of parameter space over which the emulator performs well. However if the goal is simply to perform accurate inference, we only need accurate emulation in the region of high posterior values. Motivated by this observation, in this paper we present a  `local' emulator where we iteratively query training samples from regions of high posterior. Doing so can lead to more accurate emulation in the relevant regions of the parameter space. In high dimensions, the volume of high posterior may be very small compared to the prior volume. Therefore, such active sampling schemes are crucial as we move towards high-dimensional inference that are neccessary for future surveys with increased complexity of theoretical models. 

In this paper we consider two variants of the emulator - one using Gaussian process regression and the other using neural network. We use a simulated 3x2 point analysis of LSST-Y1 to showcase the performance of our emulator. Using this case study, we demonstrate that our method leads to accurate inference while being an order of magnitude faster than standard methods. We note that iterative emulation has previously been used for cosmological analyses in \citet{PellejeroIbanez2020, Neveux2022}. In these studies, a Gaussian process emulator was used to iteratively emulate the likelihood surface. In contrast to these papers, we use our emulator to directly emulate the data vectors. For discussions on a similar likelihood acceleration tool that was developed independently from ours, we refer the reader to \cite{To2022}.

Once trained, an emulator can be used to study the impact of a wide range of systematic effects by rapidly analyzing systematics contaminated synthetic data. Further, all simulated analyses to optimize scale cuts can re-use the pretrained emulator. We note that a new emulator needs to be created for different science cases, probe combinations, galaxy samples, etc, but within these choices optimization can happen rapidly. To demonstrate this capability, we use our emulator to study the mitigation of baryonic effects in 3x2 point analyses where we achieve $\mO(1000)$ times acceleration by re-using our emulator.

We note that emulation is only necessary for parameters that enter non-trivially in the calculation of the summary statistic. As we will see later, linear galaxy bias and shear calibration calculations do not need to be emulated since their effects on the data vector can be computed instantly. This reduces the parameter space of the emulator which will be even more important for more complex physics/systematics models than we consider in this paper.

Our paper is structured as the following: in section \ref{sec:emulator}, we introduce our method for accelerating MCMC analyses. In section \ref{sec:lsst_y1}, we discuss the details of 3x2 point analysis for LSST-Y1, which is used as a case study to demonstrate our method. After presenting our results in section \ref{sec:results}, we conclude in section \ref{sec:conclusion}. We present additional details of the emulator design and Gaussian Process regression in appendix \ref{app:emu_details} and \ref{app:gp}.

For the computation of real space data vectors and covariances in this paper we use the {\sc cocoa} ({\sc cobaya}-{\sc cosmolike} {\sc architecture})\footnote{\url{https://github.com/CosmoLike/cocoa}} and {\sc CosmoCov}\footnote{\url{https://github.com/CosmoLike/CosmoCov}} codes. The former is a recently developed integration of {\sc cosmolike} \citep{Krause2017} into the {\sc cobaya} \citep{Torrado2021} framework further described in section \ref{ssec:cocoa} \citep[also see][]{Miranda:2020lpk}, the latter is a specific {\sc cosmolike} package optimized for 3x2 point real space covariances \citep{Fang2020}.
\section{Accelerating MCMC analyses using iterative emulators}\label{sec:emulator}

\subsection{Parameter inference in cosmology}\label{ssec:lkl_analyses}

In most current cosmological analyses, the full data set is first compressed into summary statistics such as two point correlation functions, peak statistics, etc. We can infer the parameters of interest (denoted as $\mvec{\Theta}$) by comparing the observed summary statistic ($\mvec{d}_{\text{obs}}$) to a model summary statistic [$\mvec{d}_{\text{model}}(\mvec{\Theta})$]. Specifically, we infer the values of $\mvec{\Theta}$ by sampling from the posterior which can be written using Bayes' theorem as,
\begin{equation}
    \mP(\mvec{\Theta}|\mvec{d_{\text{obs}}}) \propto \mP(\mvec{\Theta}) \mP(\mvec{d}_{\text{obs}}|\mvec{\Theta}),
\end{equation}
where, $\mP(\mvec{\Theta})$ is the prior imposed on the parameters and $\mP(\mvec{d}_{\text{obs}}|\mvec{\Theta})$ is related to the likelihood. The evaluation of the likelihood term involves the computation of the model data vector, $\mvec{d}_{\text{model}}(\mvec{\Theta})$ at each step of the sampling. For example, for the case of a Gaussian likelihood, the expression for the log-likelihood is 
\begin{align}
    \log \mL (\mvec{\Theta}) &\equiv \log \mP(\mvec{d}_{\text{obs}}|\mvec{\Theta}) \nonumber \\ 
    &= -\frac{1}{2}\chi^2(\mTheta) + \text{const},
\end{align}
where, 
\begin{equation}\label{eqn:chi_sq}
    \chi^2(\mTheta)= [\mvec{d}_{\text{model}}(\mvec{\Theta}) - \mvec{d}_{\text{obs}}]^{\text{T}}\mmat{C}^{-1} [\mvec{d}_{\text{model}}(\mvec{\Theta}) - \mvec{d}_{\text{obs}}].
\end{equation}
Here, $\mmat{C}$ is the data covariance. 

\subsection{Iterative strategy for acquiring training data}\label{ssec:iterative_emu}

In order to use an emulator for cosmological inference, we need accurate modelling of the data vectors in the high posterior region of parameter space. Therefore the training samples should have a good coverage of the high posterior region. Latin Hypercube samples are well-known for their space-filling property and have been widely used for training emulators \citep[e.g, ][]{Mootoovaloo2020}. However, in high dimensions, the posterior volume can be significantly smaller than the prior volume. In fact, the number of points required to have a fixed number of points in the high probability region goes up exponentially with dimensions.

\begin{figure}
    \centering
    \includegraphics[width=0.8\linewidth]{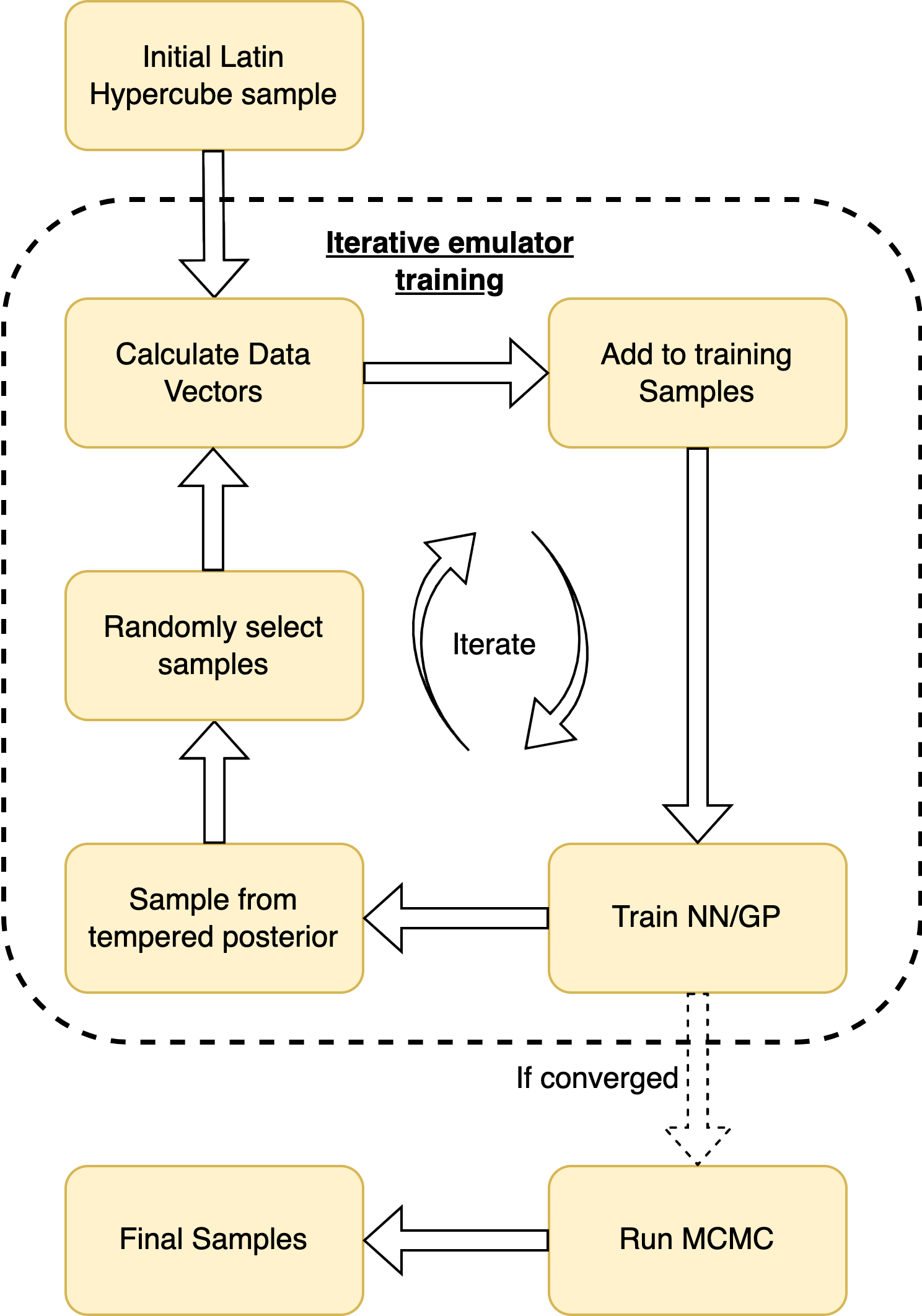}
    \caption{Schematic description of the emulator design}
    \label{fig:iterative_emulator}
\end{figure}

Fortunately, for the purposes of inference, we only need accurate emulation in the parameter range of our interest, i.e, in regions of high posterior value. Therefore, we need high density of training samples only in these regions of the parameter space. In order to achieve this, we devise an iterative emulator where we take the following steps: 
\begin{itemize}
    \item[{\it i)}] We start with a Latin Hypercube (LH) sample that covers the prior volume of the parameter space. We train our first emulator only on these LH samples to predict the model data vector as a function of the parameters, $\mvec{d}_{\text{emu}}(\mvec{\Theta})$. We calculate the $\chi^2$ for all training points and discard the training points with $\chi^2 > \chi^2_{\text{cutoff}}$.
    \item[{\it ii)}] We acquire more training data in the vicinity of high posterior region using the emulator. To do so, we run an MCMC analysis with the tempered posterior, where the likelihood is artificially inflated by a tempering factor, $\alpha < 1$, such that 
    \begin{equation}
        \log \mL_{\text{tempered}}(\mvec{\Theta}) = \alpha \times \log \mL (\mvec{\Theta}).
    \end{equation}
    The sampling step is now fast since we rely on the emulator for the sampling. We then add $N_{\text{train}}$ randomly selected  samples from the tempered posterior to our training data set.
    \item[{\it iii)}] We calculate the model data vectors for the newly acquired training samples and re-train our emulator. For our NN emulator, we train our neural network on the combined sample of all the training samples until that iteration. For the Gaussian process, we only train it on the training data set from that iteration.
    \item[{\it iv)}] We iterate steps {\it (ii)-(iii)} for $N_{\text{iter}}$ iterations. In this paper, we use $N_{\text{iter}}=4$. We can also define a convergence criteria to select $N_{\text{iter}}$. But we do use such criteria in this paper.
    \item[{\it v)}] Once the iterative training is done, we use the trained emulator to run our MCMC analysis to get samples from the posteriors.
\end{itemize}
 The details of the emulator is presented in section \ref{ssec:emu_design}. We illustrate the design of our iterative emulator scheme in Fig. \ref{fig:iterative_emulator}. We use the publicly available {\sc python} package {\sc emcee}\footnote{\url{https://emcee.readthedocs.io/}} \citep{ForemanMackey2013} to sample from the emulator posteriors.

\subsection{Emulator design}\label{ssec:emu_design}

In our emulator, we predict the model data vector, $\mvec{d}_{\text{model}}$ as a function of parameters, $\mvec{\Theta}$. These parameters include cosmological as well as systematic parameters. The resulting dimensions of the parameter space for the analysis of Stage-IV surveys can be fairly high  ($\gtrsim 50$). Therefore our emulation method needs to scale with dimensions. Among the different emulation methods, Gaussian process (GP) regression and neural networks (NN) scale well with dimensions and have complementary advantages and drawbacks. We implement both these types of  emulators in this work, thus utilizing the complementary advantages of the two methods and providing a test of robustness of the inferred parameters. 
The details of the GP and NN emulators are presented in sections \ref{sssec:gp_emu} and \ref{sssec:nn_emu} respectively. Further details of the emulator design is presented in Appendix \ref{app:emu_details}.
\subsubsection{Gaussian Process emulator}\label{sssec:gp_emu}

The first type of emulator we use in this work uses Gaussian process (GP) regression \citep{Rasmussen2006} to predict the value of the data vectors. GP regression is a non-parametric Bayesian method that predicts the value of a function conditioned on a set of training data. Gaussian process regression scales well with dimensions making it an ideal candidate for our high-dimensional emulator. On the flip side, GP regression does not scale well with the number of training data points. The technical details of GP regression is presented in Appendix \ref{app:gp}.

We use the {\sc python} package {\sc george}\footnote{\url{https://george.readthedocs.io/}} \citep{Ambikasaran2014} to implement our Gaussian process emulator. We use separate single output GP regressors to predict the value of each element of the data vector. That is, the $i$-th element of the data vector is predicted at some parameter value $\mTheta$ using the $i$-th Gaussian process regressor, $\text{GP}^{i}$ as,
\begin{equation}
    \mvec{d}^{i}_{\text{model}}(\mTheta) = \text{GP}^{i}(\mTheta|\{\mTheta_{\text{train}}\}, \{\mvec{d}^i_{\text{train}}\}),
\end{equation}
where, $\{\mTheta_{\text{train}}\}$ and $\{\mvec{d}_{\text{train}}\}$ are the training data. Thus, we use a total of $N_{\text{dv}}$ GPs to emulate the data vector, where $N_{\text{dv}}$ is the number of elements in the data vector.

\subsubsection{Neural Network emulator}\label{sssec:nn_emu}
We also use a neural network emulator in this work. As opposed to Gaussian process regression, neural network scales well with training data, i.e, it performs better with a larger training data. Therefore, given that the network is sufficiently expressive, we can improve the performance of the emulator by simply adding more training data.

We use a simply connected neural network with $3$ hidden layers with $1024$ neurons each to predict the data vectors as a function of the $\Theta$. Each layer is linearly connected to all the neurons of the previous layer, followed by the application of a Rectified Linear Unit (ReLU) activation function. We use a $L_1$ loss,
\begin{equation}
    L(\mvec{w}) = \sum_{\mTheta \in \mathcal{T}} \sum_{i=1}^{N_{\text{dv}}}|\mD^{i}_{\text{NN}}(\mTheta; \mvec{w}) - \mD^i_{\text{model}}(\mTheta)|,
\end{equation}
 to train the neural network. In the above, $\mathcal{T}$ denotes the sample of training data, $\mvec{w}$ are the weights of the neural network that are optimized in the training, $\mD_{\text{NN}}$ is the normalised data vector computed using the neural network and $\mD_{\text{model}}$ is the data vector computed using the full model. Here, the data vectors are normalized according to equation \eqref{eqn:dv_normalization}. We use the Adam optimizer \citep{Kingma2014} to optimize the weights of the neural network. We implement our neural network emulator in {\sc pytorch} \citep{pytorch}.
\section{Case Study: LSST-Y1 3x2 analysis}\label{sec:lsst_y1}

\begin{table}
  \centering
  \caption{Fiducial value and the prior used on various cosmological and systematic parameters. In the table flat$[a,b]$ denotes a flat prior between $a$ and $b$, whereas Gauss$[\mu, \sigma^2]$ denotes a Gaussian prior with mean $\mu$ and standard deviation $\sigma$. See section \ref{ssec:systematics} for more detail on the modelling of the systematic parameters.}
  \begin{tabular}{l c c}
  \hline
     Parameters &  Fiducial & Prior\\
    \hline
    \textbf{Survey} & & \\
    Area, $\Omega_s$ & 12300 deg$^2$ & fixed\\
    Shape noise, $\sigma_e$ & $0.26$& fixed\\
    \hline
    \textbf{Cosmology} & & \\
    $\Omega_\textrm{m}$ & $0.3$ & flat$[0.01, 0.9]$\\
    $\log (A_\textrm{s} \times 10^{10})$ & $3.0675$ & flat$[1.61, 3.91]$\\
    $n_{\textrm{s}}$ & $0.97$ & flat$[0.87, 1.07]$\\
    $\Omega_\textrm{b}$ & $0.048$& flat$[0.03, 0.07]$\\
    $h$ & $0.69$& flat$[0.55, 0.91]$\\
    \hline
    \textbf{Intrinsic Alignment} & & \\
    $a_{\ia}$ & $0.5$ & flat$[-5, 5]$\\
    $\eta_{\ia}$ & $0$ & flat$[-5, 5]$\\
    \hline
    \textbf{Linear galaxy bias} & & \\
    $b^{(i)}_1$& $[1.24, 1.36, 1.47$ & flat$[0.8, 3]$ \\
    & $1.60, 1.76]$&\\
    \hline
    \textbf{Photo-$z$ bias} & & \\
    $\Delta^i_{z,\text{lens}}$ & 0 & Gauss$(0., 0.005^2)$\\
    $\Delta^i_{z,\text{source}}$ & 0 & Gauss$(0., 0.002^2)$\\
    \hline
    \textbf{Shear calibration} & & \\
    $m^i$ & 0 & Gauss$(0., 0.005^2)$\\
    \hline
    \textbf{Baryon PCA amplitude} & & \\
    $Q_1$ & 0 & flat$[-3, 12]$\\
    $Q_2$ & 0 & flat$[-2.5, 2.5]$\\
    $Q_{> 2}$ & 0 & fixed\\
    \hline
  \end{tabular}
  \label{tbl:parameters}
\end{table}

In order to demonstrate the performance of our emulator, we present a realistic forecast of 3x2 point analysis for an LSST-Y1 like survey. 3x2 point analysis refers to the combined cosmological analysis using three 2 point functions: cosmic shear ($\xi_{\pm}$), galaxy-galaxy lensing (GGL, $\gamma_t$), and galaxy clustering ($w$). In section \ref{ssec:survey}, we present the details of the assumed survey configuration, tomographic binning and scale cuts. As detailed in section \ref{ssec:systematics}, we also model various systematic effects such as the intrinsic alignment, photo-$z$ bias, galaxy bias, multiplicative shear calibration bias and baryonic effects. In this work, we analyze the real space angular correlation functions following closely the analysis of \citet{Fang2020}. The parameters used in our likelihood analysis along with the prior imposed on these parameters are summarized in Table \ref{tbl:parameters}.

\subsection{Survey assumptions and analysis choices}\label{ssec:survey}

The survey assumptions for LSST-Y1 survey is based on the DESC Science Requirement Document \citep[SRD, ][]{LSSTDESC_SRD}. The mock LSST-Y1 survey has a survey area of $12300$ deg$^2$. The shape noise per component is assumed to be $\sigma_{e} = 0.26$. The redshift distribution for the source and lens samples are computed assuming a Smail distribution \citep{Smail1995}, where $n(z) \propto z^2 \exp[-(z/z_0)^{\alpha}]$. 
For the source sample, the assumed value of the Smail distribution parameters are $(z_0, \alpha) = (0.191, 0.870)$  \citep{Fang2020}. The effective number density of source galaxies is $n_{\text{eff}} = 11.2$ arcmin$^{-2}$.
The distribution of the lens galaxies are parameterized with the values $(z_0, \alpha) = (0.26, 0.94)$ with an effective number density, $n_{\text{eff}} = 18$ arcmin$^{-2}$. The lens and the source samples are each divided into 5 tomographic bins with equal number of galaxies and each tomographic bin is convolved with a Gaussian photo-$z$ uncertainty of $0.02 (1 + z)$ and $0.05 (1+z)$ respectively. 

We compute the angular correlation function in $26$ logarithmically spaced angular bins between $2.5^{\prime}$ and $900^{\prime}$. Without any scale cuts, the full data vector, consisting of $\{\xi_{+}, \xi_{-}, \gamma_t, w\}$, has a total of $1560$ elements. Since the theoretical modelling can break down at small scales due to insufficient accuracy in the theoretical modelling, we impose an angular scale cut for the 2-point functions. Following \citet{Fang2020}, we only use angular scales corresponding to physical scales larger than $R_{\text{min}} = 21~h^{-1}$ Mpc for galaxy clustering and galaxy-galaxy lensing parts of the data vector. This corresponds to a angular scale cuts of $\theta > [80.88^{'}, 54.19^{'}, 42.85^{'}, 35.43^{'}, 29.73^{'}]$ for the $5$ tomographic bins. 

For cosmic shear, \citet{Fang2020} imposed a scale cut corresponding to $\ell_{\text{max}} = 3000$. In this paper, we model baryonic physics using the PCA marginalization approach of \citet{Eifler2015, Huang2019}, which allows us to use all scales in the cosmic shear part of the data vector with tolerable biases even for extreme feedback scenarios (see section \ref{ssec:baryon_scale_cuts}). The simulated observed data vectors are produced with our theoretical model at the fiducial parameter values given in Table \ref{tbl:parameters}.

We use an analytic covariance, $\mmat{C}$, for the data vectors. The covariance is computed using the \cosmolike \citep{Krause2017} software. \cosmolike calculates the covariance using an analytical approach and calculates the Gaussian, connected non-Gaussian and the super-sample part of the covariance.

\subsection{Systematics modelling}\label{ssec:systematics}

In this section, we outline the modelling of different systematic effects in our theoretical data vectors.

\textbf{Intrinsic alignment}: We use the non-linear alignment (NLA) model \citep{Bridle2007} to model intrinsic alignment in our data vectors. In the NLA model, the redshift dependent amplitude of the IA signal is modelled as,
\begin{equation}
    A_{\text{IA}}(z) = -a_{\ia}\frac{C_1 \rho_{\text{cr}} \Omega_m}{G(z)}\bigg(\frac{1+z}{1+z_0}\bigg)^{\eta_{\ia}},
\end{equation}
where, $C_1 \rho_{\text{cr}} = 0.0134$, $z_0 = 0.62$, $G(z)$ is the linear growth factor, $a_{\ia}, \eta_{\ia}$ are the parameters of the NLA model parameterizing the amplitude and the redshift dependence of the IA signal respectively. We use a flat prior between $-5$ and $5$ for $a_{\ia}$ and  $\eta_{\ia}$.

\textbf{Galaxy bias}: We use a linear galaxy bias model to model the galaxy bias of the lens sample. Linear galaxy bias is a fast parameter (see section \ref{ssec:fast_parameters}) and modifies the data vectors in a simple multiplicative manner. Therefore, we fix the value of galaxy bias to a fiducial value, $b_{1, \text{fid}}$ while computing the training data vectors for the emulator. To model its impact, the galaxy-galaxy lensing part, $\gamma_t$, and the galaxy clustering part, $w$, part of the emulated data vectors are then modified as,
\begin{align}
    \gamma_t &\rightarrow \frac{b_1}{b_{1,\text{fid}}} \times \gamma_t,\\
    w &\rightarrow \bigg(\frac{b_1}{b_{1,\text{fid}}}\bigg)^2 \times w.
\end{align}
We impose a flat prior between $[0.8, 3]$ on each of the $5$ galaxy bias parameters.

\textbf{Photo-$z$ bias}: The uncertainty in the redshift distribution of galaxies in photometric surveys is modelled and marginalizing using photo-$z$ bias parameter, $\Delta^i_z$, which shifts the estimated redshift distribution, $\hat{n}$ as,
\begin{equation}
    n^i(z) \rightarrow \hat{n}^i(z + [1 + \bar{z}^i] \Delta^i_z),
\end{equation}
where, $\bar{z}^i$ is the mean redshift of the $i$-th tomographic bin. 

We assume a Gaussian prior centred on $0$ and a standard deviation of $0.002$ ($0.005$) for each of the 5 source (lens) bin. We use a total of $10$ parameters to characterize the photo-$z$ biases.

\textbf{Multiplicative shear calibration bias}: We use one multiplicative shear calibration bias parameter per each of the $5$ source bin. Shear calibration parameters are also fast multiplicative parameters. While calculating the training data vectors, the value of the shear calibration biases are set to zero. During the inference, its impact on the cosmic shear and the galaxy-galaxy lensing part of the data vectors is modelled as,

\begin{align}
    &\xi^{ij}_{\pm}(\theta) \rightarrow (1 + m^i)(1 + m^j) \xi^{ij}_{\pm}(\theta; m=0), \\
    &\gamma^{ij}_{t}(\theta) \rightarrow (1 + m^i) \gamma^{ij}_{t}(\theta; m=0).
\end{align}
We impose a Gaussian prior centred on $0$ with a standard deviation $0.005$ on each of the $5$ shear calibration parameters.

\textbf{Baryonic effects}: Baryonic feedback can substantially impact the small scale power spectrum and therefore can lead to biased inference of cosmological parameters. To mitigate the impact of baryonic effects in the data vectors, we use the principal component analysis (PCA) marginalization approach of \citet{Eifler2015}. We model the baryonic feedback only for the cosmic shear ($\xi_{\pm}$) part of the data vector. 
In the PCA marginalization approach: 
\begin{itemize}
    \item First, the deviations in the data vector, $\Delta \mvec{d}$, due to baryonic effects are calculated from $N_{\text{sim}}$ hydrodynamic simulations. In this paper, we use the data vector deviations from a total of $8$ hydrodynamical simulations:
    Illustris TNG-100 \citep{Springel2018}, MassiveBlack-II \citep{Khandai2015}, Horizon-AGN \citep{Dubois2014}, 2 simulations (T8.0 and T8.5) of the cosmo-OWLS simulation suite \citep{LeBrun2014} and 3 simulations (T7.6, T7.8, T8.0) of the BAHAMAS simulation suite \citep{McCarthy2017}.
    \item A PCA decomposition is performed for the data vector deviations to compute a set of principal components, $\{\text{PC}_n\}$. We refer to \citet{Eifler2015, Huang2019, Huang2021} for more details.
    \item In our likelihood analysis, the amplitude of the PCA modes, $Q_n$, are treated as free parameters that can be inferred from the data. For an LSST like survey, \citet{Huang2019} found that 2 PCA modes are sufficient to mitigate the baryonic effects for most feedback scenarios.
    Therefore, we only use the first two PC amplitudes, $Q_{1/2}$ to model the baryonic feedback and set all the higher PC amplitudes, $Q_{>2}$ to $0$. The model data vector after accounting for the baryons is given as,
    \begin{equation}
        \mvec{d}_{\text{model}} = \mvec{d}_{\text{dmo}} + \sum_{n=1}^{2} Q_n \text{PC}_n,
    \end{equation}
    where $\mvec{d}_{\text{dmo}}$ is the model data vector with no baryons. Since the impact of the baryons can be modelled simply by addition of two PCA modes, it is a fast parameter that is not used in the emulation.
\end{itemize}

\subsection{Data vector computation and MCMC sampling}
\label{ssec:cocoa}
We use the newly developed {\sc cobaya}-{\sc cosmolike} {\sc architecture} ({\sc cocoa}), to calculate the theoretical data vectors as a function of  cosmological and systematic parameters. {\sc cocoa} is a code framework that integrates \cosmolike \citep{Krause2017},  into the \cobaya likelihood framework \citep{Torrado2021}. 

The new implementation improves previous {\sc cosmolike} versions in two important aspects: the code can be threaded with shared memory parallel programming (e.g., OpenMP) and further parallelized using the Message Passing Interface (MPI) standard; it can now can receive distances and power spectrum tables from any Boltzmann code (or emulator), e.g.,  {\sc class} Boltzmann code \citep{Blas2011} and \camb \citep{Lewis2000, Howlett2012}. Extensive testing of this pipeline has been presented in appendix A of \citet{Miranda:2020lpk}.

In this paper we adopt \camb to calculate the matter power spectrum, and Takahashi Halofit \citep{Takahashi2012} to compute non-linear power spectrum corrections. The calculation of projected power spectra and the calculation of systematic effects closely follows the {\sc cosmolike} implementation described in \citep{Fang2020}.

To perform the full likelihood analysis without the use of the emulator, we use the slow-fast sampler used in {\sc cobaya} \citep{Lewis2002, Lewis2013}. 
\section{Results}\label{sec:results}

\begin{figure*}
    \centering
    \includegraphics[width=\linewidth]{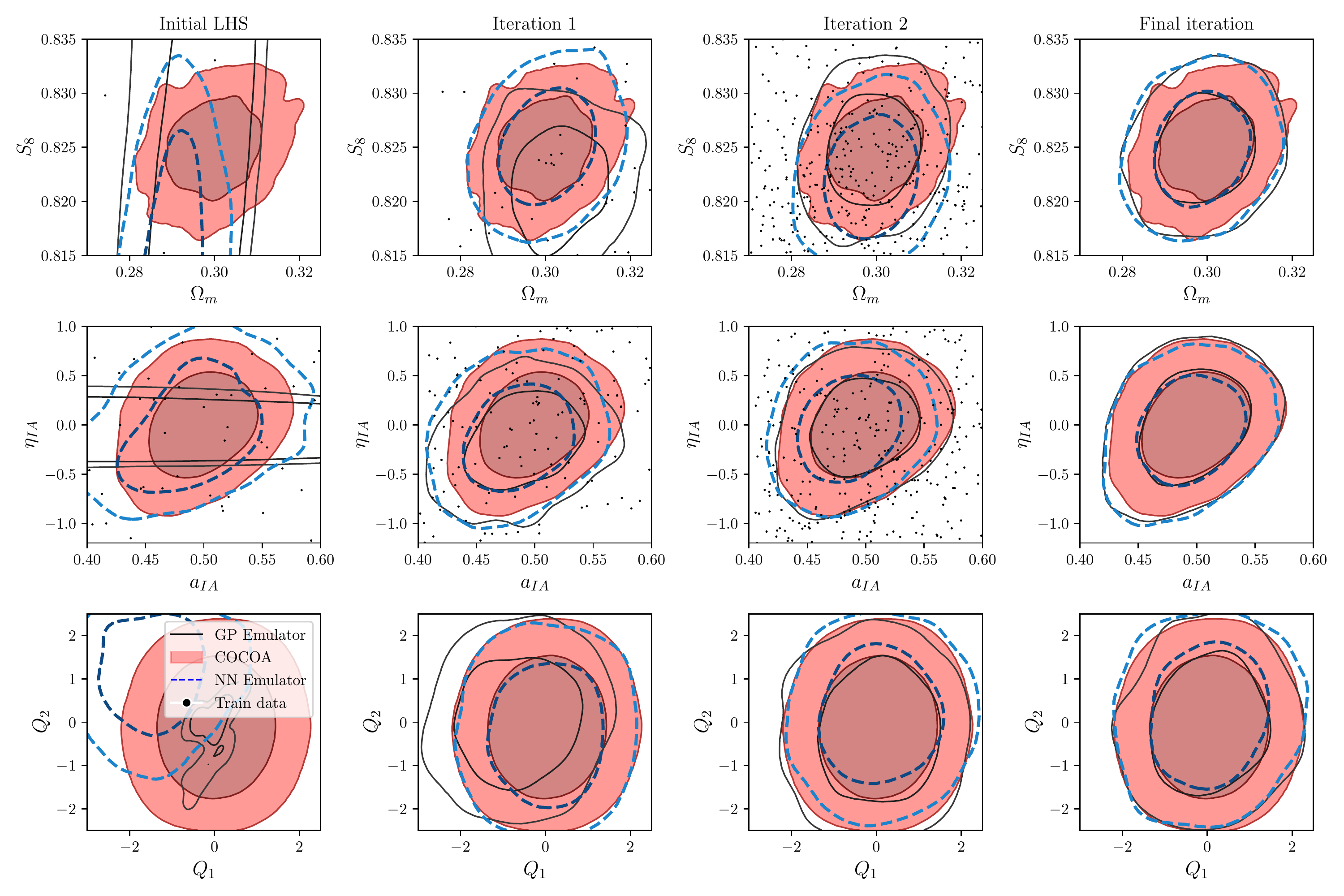}
    \caption{The contours in the $\Omega_m$-$S_8$ plane ({\it top}), intrinsic alignment parameters, $a_{\text{IA}}$-$\eta_{\text{IA}}$ ({\it middle}) and the baryon PCA parameters, $Q_1$-$Q_2$ ({\it bottom}). The different columns correspond to the different iterations of emulator training. The red shaded contours are produced with the full theoretical model of {\sc cocoa}, black contours are produced with our GP emulator and the blue-dashed contours are from the NN emulator. The black points in the top and middle row show the training points sampled from the tempered posterior in that iteration. Since we are not emulating the $Q$ parameters, there are no training points in this plane. The contours produced by the emulators improves towards the later iterations. From the rightmost column, we can see that both the NN emulator as well as the GP emulator is successful in reproducing the {\sc cocoa} contours with high accuracy up to sampling noise. }
    \label{fig:training_iteration}
\end{figure*}

In this section we present the results of our emulator based 3x2 point analysis of an LSST-Y1 like survey. In section \ref{ssec:iter_emu}, we compare the cosmological constraints based on our emulators to a full theoretical model based likelihood analysis using {\sc cocoa}. In section \ref{ssec:emu_accuracy}, we discuss the accuracy of our emulator. Section \ref{ssec:gp_vs_nn} compares the performance of the Gaussian process and the neural network emulator. In section \ref{ssec:importance_sampling}, we discuss a resampling strategy using importance sampling that is guaranteed to recover the correct posterior. Finally, in section \ref{ssec:baryon_scale_cuts}, we use our trained emulator to run extremely quick analyses to study different scale cuts and strategies to mitigate baryonic effects in LSST-Y1 3x2 point analyses. 

\subsection{Cosmological inference using an iterative emulator}\label{ssec:iter_emu}

\begin{figure}
    \centering
    \includegraphics[width=\linewidth]{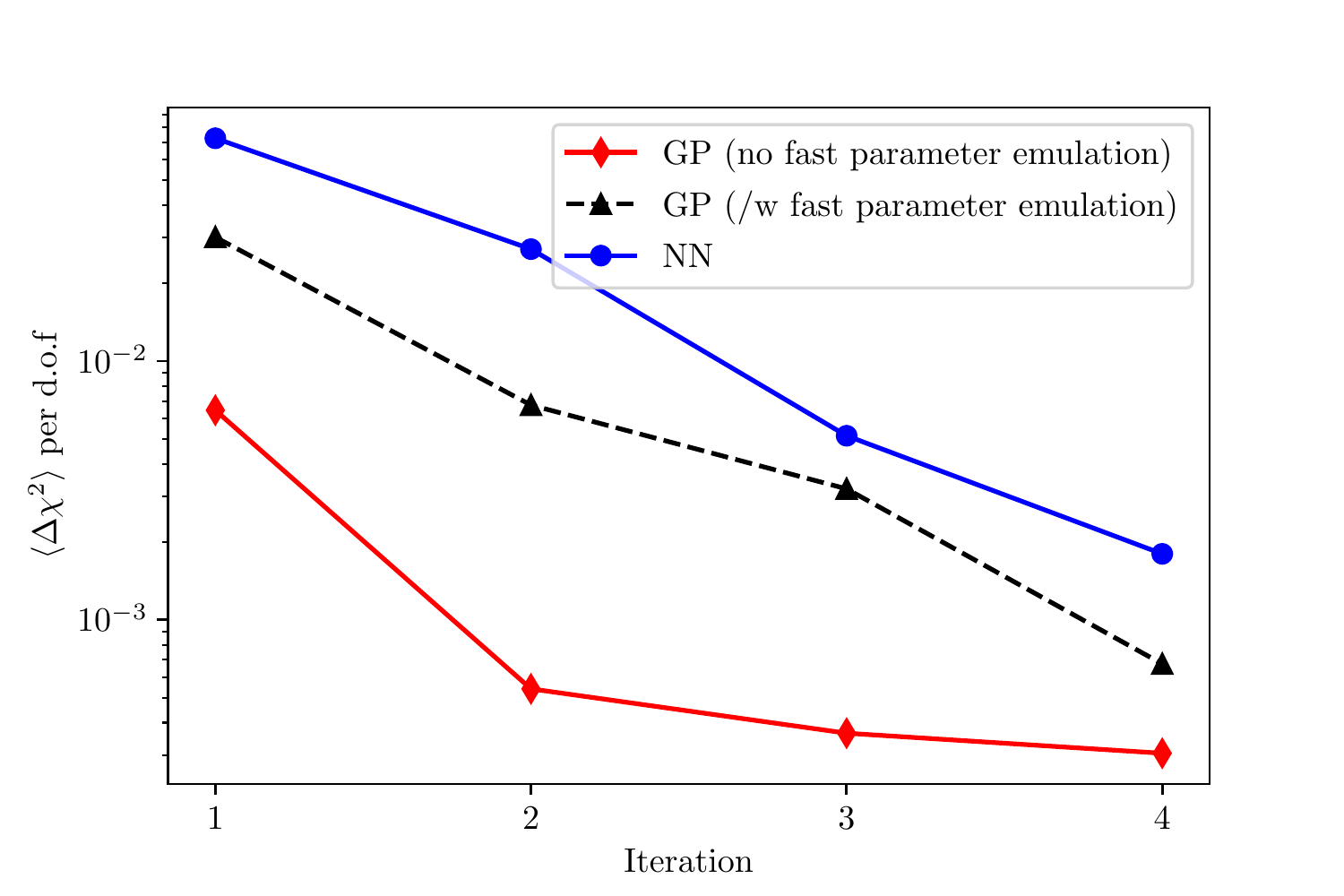}
    \caption{The shift in $\chi^2$ per d.o.f averaged over a test set at each iteration of the emulator training. The red diamonds show $\Delta \chi^2$ per d.o.f for the GP emulator, while the blue circles show the same quantity for the NN emulator. The black dashed triangles show the shift for a GP emulator where we emulate all parameters. The shift in $\chi^2$ per d.o.f is insignificant showing that our emulators leads to accurate results. A comparison of the different curves indicates that our GP emulator performs slightly better than the NN emulator (See section \ref{ssec:gp_vs_nn}). Furthermore, we see that emulator accuracy increases if we only emulate the non-linear parameters.}
    \label{fig:delta_chi_sq}
\end{figure}

As described in section \ref{ssec:iterative_emu}, our emulator involves iterative sampling, meaning we add training samples from a tempered posterior in each iteration. The initial training sample is drawn from a Latin Hypercube (LH) sampling of the prior range. For parameters with Gaussian priors (e.g, photo-$z$ biases), we use a LH sample between $-3\sigma$ to $+3\sigma$ of the prior. In the subsequent iterations, we add training samples by sampling from a tempered posterior. 

We initialize our run with a total of 9600 Latin Hypercube samples and select training points with  $\chi^2 < 10^4$. After training the emulator on this set of training data, we sample from a tempered posterior using the emulator. We start with a small value of $\alpha$ and then progressively increase it. The values of $\alpha$ used in our training for the $4$ iterations are $[0.05, 0.1, 0.3, 0.5]$. From these sampled points, we then randomly select $960~(4800)$ points for our next iteration of training sample for the GP (NN) emulator. 

We show the iterative sampling and the addition of the training points in Fig. \ref{fig:training_iteration}. We show the contours derived from the full \cocoa model along with the emulator contours and the training samples added for each iteration. We show this for three sets of parameters: {\it i)} Cosmological contours, $\Omega_m$-$S_8$, {\it ii)} Intrinsic alignment parameters, $a_{\text{IA}}$-$\eta_{\text{IA}}$ and {\it iii)} Baryon PC amplitudes, $Q_1$-$Q_2$. Since the baryon PC amplitudes are fast parameters, they are not used in the emulator and therefore we do not plot any training samples in those parameters. 

The emulator-based contours improve progressively towards the later iterations. Furthermore, the density of the training points in the region of high posterior value increases. This is because we reduce the temperature of the tempered likelihood for the later iterations. Finally, we also note that both our GP-based emulator and the NN-based emulator are able to reproduce the \cocoa contours with high accuracy. 

\subsection{Emulator accuracy}\label{ssec:emu_accuracy}

In this section, we assess the accuracy of our emulators.
We calculate the shift in $\chi^2$ per degrees of freedom due to error in emulation.
To calculate this quantity, we create a test set by drawing $1000$ randomly selected points from the \cocoa MCMC chain. For this test set, we calculate the model data vectors using \cocoa and then calculate the average shift in $\chi^2$ per d.o.f from the test set,
\begin{equation}
    \langle \Delta \chi^2 \rangle \text{/ d.o.f} = \frac{1}{N_{\text{test}}}\sum_{i=1}^{N_{\text{test}}}\frac{1}{N_{\text{dof}}}[\chi^2_{\text{emulator}}(\mTheta_i) - \chi^2_{\text{cocoa}}(\mTheta_i)].
\end{equation}
%
We show the average shift in $\Delta\chi^2$ per d.o.f in the test set in Fig. \ref{fig:delta_chi_sq} for our emulators at each training iteration. In that figure, along with the comparison of the GP and the NN emulator, we also show the results of an emulator strategy where we use a GP emulator that emulates the fast parameters. 
The average shift in $\chi^2$ per d.o.f calculated for the test set calculated using the trained GP emulator is $0.00035 \pm 0.00166$. The same shift for the NN emulator is $0.00253 \pm 0.00276$. These shifts in $\chi^2$ per d.o.f are lower than or at the level of shifts in $\chi^2$ per d.o.f due to different methods of calculating the covariance matrix \citep[see ][]{Fang2020}.

As mentioned in section \ref{ssec:fast_parameters} we do not emulate the fast parameters in the model, but compute their impact perfectly on the emulated quantity. Specifically, for the LSST-Y1 3x2 point analysis, we do not emulate the linear galaxy bias parameters, $b^{i}_1$, shear calibration parameters, $m_i$, and the Baryon PC amplitudes, $Q_{1/2}$. The impact of this choice can be seen in Fig. \ref{fig:delta_chi_sq}, where we compare GP emulators with (black dashed triangles) and without fast parameters (red diamonds). The shift in $\chi^2$ per d.o.f when emulating the full parameter space is $0.00058 \pm 0.00123$ (compared to $0.00035 \pm 0.00166$ when emulating the non-linear subspace). While both GP emulators leads to insignificant shifts in $\chi^2$ per d.o.f., we note that this can change when more complex parameter spaces and parameter dependencies are considered. Given that the impact of the fast parameters can be calculated exactly, emulating only the subspace of non-linear parameters is the preferred method.  

\subsection{Comparison of the GP and NN emulators}\label{ssec:gp_vs_nn}

We now compare the performance and other aspects of the GP and NN emulators used in this work. As we already saw in the previous sections, both the GP and NN emulators reproduce the correct cosmological contours. While we find some indication that the GP emulator performs slightly better than the NN emulator, this statement is likely subject to change, e.g., when using different architectures for the NN or when considering different parameter spaces. 

The GP and the NN regression are highly complementary. After training, the neural network sampling is extremely fast. Compared to the GP, neural networks may require a large training data set and tuning of neural network architecture and hyper-parameters for optimal accuracy. In contrast, GP regression is generally more robust to the details of training but it does not scale well with a large training data set. Furthermore, the prediction of the GP emulator can be much slower than a NN emulator. 

\begin{figure}
    \centering
    \includegraphics[width=\linewidth]{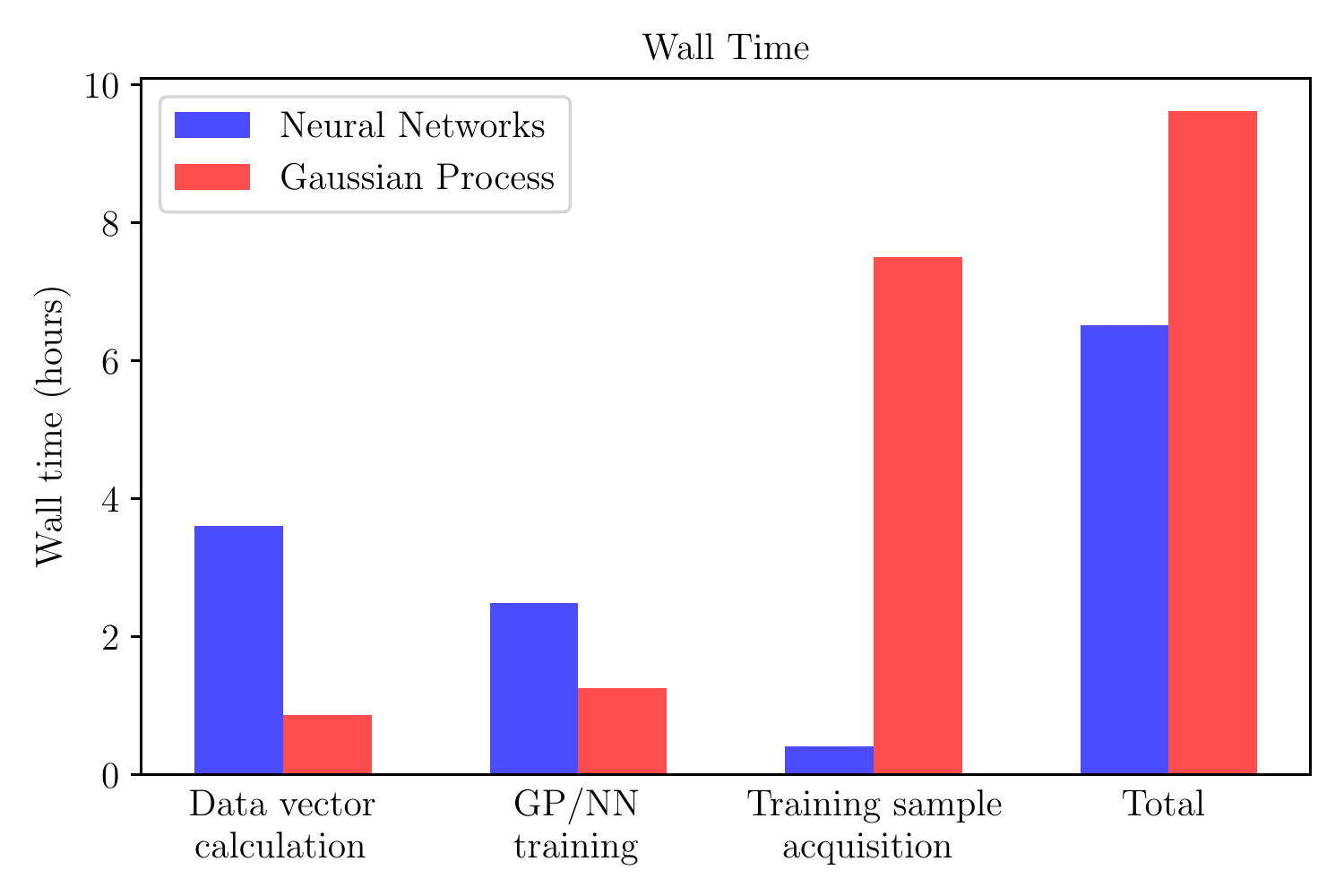}
    \caption{Breakdown of the wall time required for our analysis with the GP (red) and NN emulators (blue) showing the time required for calculation of the data vectors for the training sample, training the emulator, acquisition of the training sample from the tempered posterior and the total time required. We see that the main computational bottleneck for the NN emulator is the calculation of the large number of data vectors. On the other hand, the main bottleneck for the GP emulator is the acquisition of the training samples from the tempered posterior.}
    \label{fig:timing}
\end{figure}

In Fig. \ref{fig:timing}, we show the wall time required for our analysis with both the Gaussian Process and the Neural network emulator on a 48 CPU core computer. The total wall time required for the analysis using the NN and the GP emulators are $\sim 5$ hours and $\sim 8$ hours respectively. For comparison, the MCMC run on the same computer with the full theoretical model took $\sim 100$ hours of wall time. For our configuration, the use of the iterative emulator therefore leads to a speed-up of an order of magnitude. 

Note that the breakdown of the timing for the two emulators are very different. The main computational expense for the NN emulator is the computation of the large number of data vectors and training the neural network. On the other hand, GP does not require a large training set of data vectors. Instead, time required for each model evaluation with the GP emulator is much larger than that with the neural network. Consequently, the acquisition of new training samples is the computational bottleneck for our GP emulator. 

As always, wall time numbers are machine and implementation dependent. It is important to note that the calculation of the data vectors, the main bottleneck for the NN, can be fully parallelized. 

\subsection{Importance sampling of the approximate posterior}\label{ssec:importance_sampling}

So far, we have derived the cosmological parameters iteratively such that the derived contours improve at each iterations. In this section, we present a way to further improve the derived cosmological contours using importance sampling. Importance sampling is a method that allows us to infer properties of a distribution, $\mP(\mTheta)$, by sampling from another approximate distribution, $Q(\mTheta)$, and then assigning an importance weight to the samples drawn from $Q$ \citep[see e.g,][]{Mackay2003}.

In our iterative emulation scheme, the samples derived using the emulator can be assigned an importance weight to re-weigh the final samples. Using importance sampling in that manner is theoretically guaranteed to produce the correct contours even if the emulator is not very accurate. However, if the posterior derived from the emulator has a large bias and does not contain sufficient samples from high posterior regions of the parameter space, this weighting scheme may lead to unreliable results. To mitigate possible biases in the posterior of this nature, we sample using our emulator from a tempered posterior with tempering factor, $\alpha$, and then compute the importance weight for the samples. Sampling from the tempered posterior leads to a broader distribution that `covers' the true posterior. The importance weight for this case is calculated as,
\begin{equation}
    \mathrm{w}_{\text{I}} = \frac{\mP_{\text{true}}(\mvec{\Theta})}{\mP_{\text{emu}}(\mvec{\Theta}; \alpha)},
\end{equation}
where, $\alpha$ is the tempering factor, $\mP_{\text{emu}}$ is the probability evaluated using the emulator and $\mP_{\text{true}}$ is the probability re-evaluated by calculating the full data vector for the selected samples. Note that in this scheme, we need to evaluate the likelihood for all the samples needed for the re-weighting. 

\begin{figure}
    \centering
    \includegraphics[width=\linewidth]{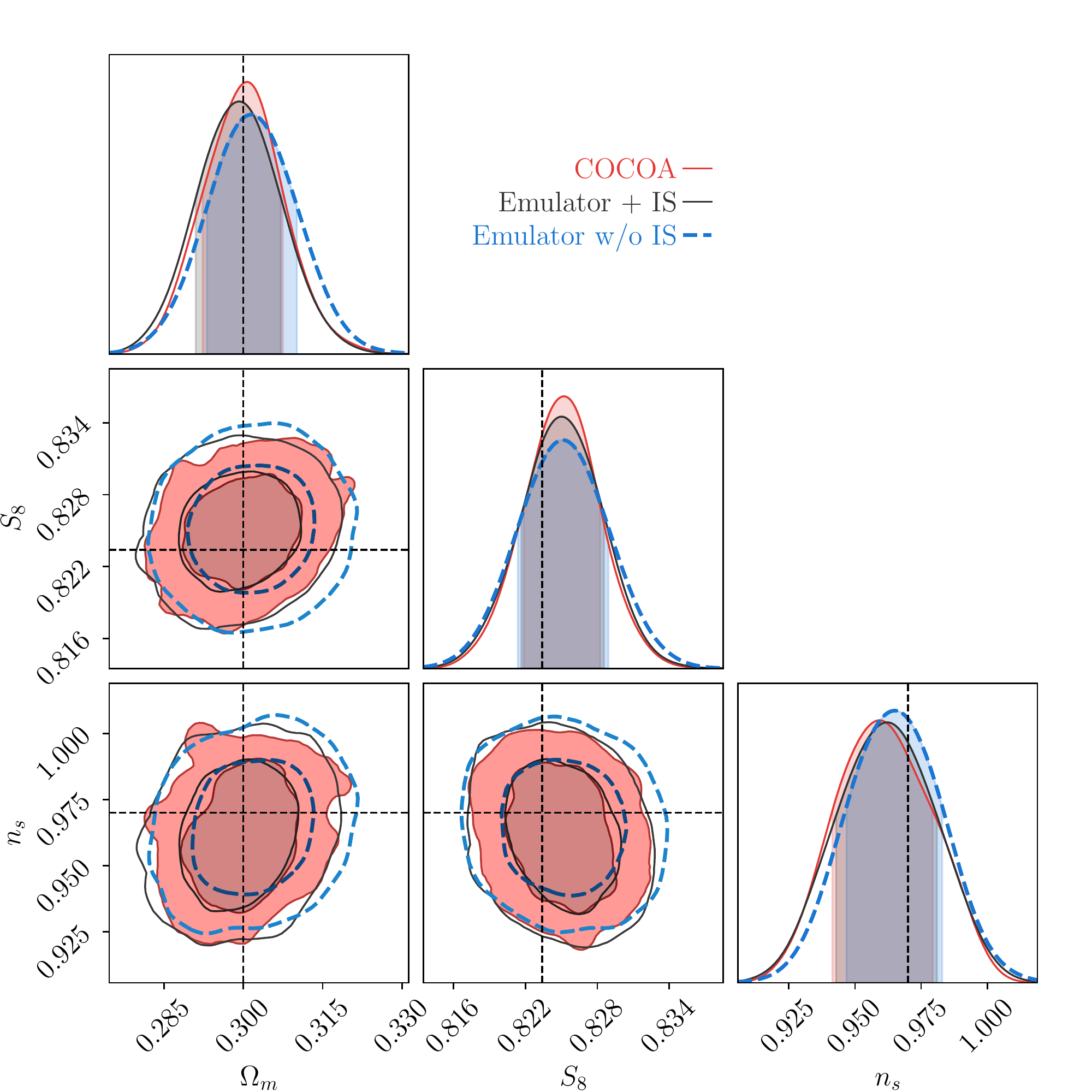}
    \caption{Cosmological parameters derived using the full theoretical model (red shaded contours), the iterative emulator without importance sampling (blue dashed contours) and using the emulator followed by importance sampling (black contours). The black dashed lines show the true fiducial parameters. The importance sampling scheme is theoretically guaranteed to produce accurate contours even in the presence of emulator inaccuracies. In this case, the emulator produces highly accurate posteriors even without importance sampling. For more details, refer to section \ref{ssec:importance_sampling}.}
    \label{fig:importance_sampling}
\end{figure}

We show the cosmological contours derived using this strategy in Fig. \ref{fig:importance_sampling} along with the contours derived using the full theoretical model with \cocoa and the contours derived using the iterative emulator but without importance sampling. For the demonstration in this section, we train our NN emulator for 3 iterations and then sample using that emulator from a tempered posterior with $\alpha=0.7$ before using importance sampling to derive the final posterior. We use a total of $96000$ samples for the importance sampling. Our iterative emulator already produces highly accurate cosmological contours (see section \ref{ssec:iter_emu}) and we do not see any major improvement on using the importance sampling scheme. 

\subsection{Using the emulator to determine scale cuts}\label{ssec:baryon_scale_cuts}

\begin{figure}
    \centering
    \includegraphics[width=\linewidth]{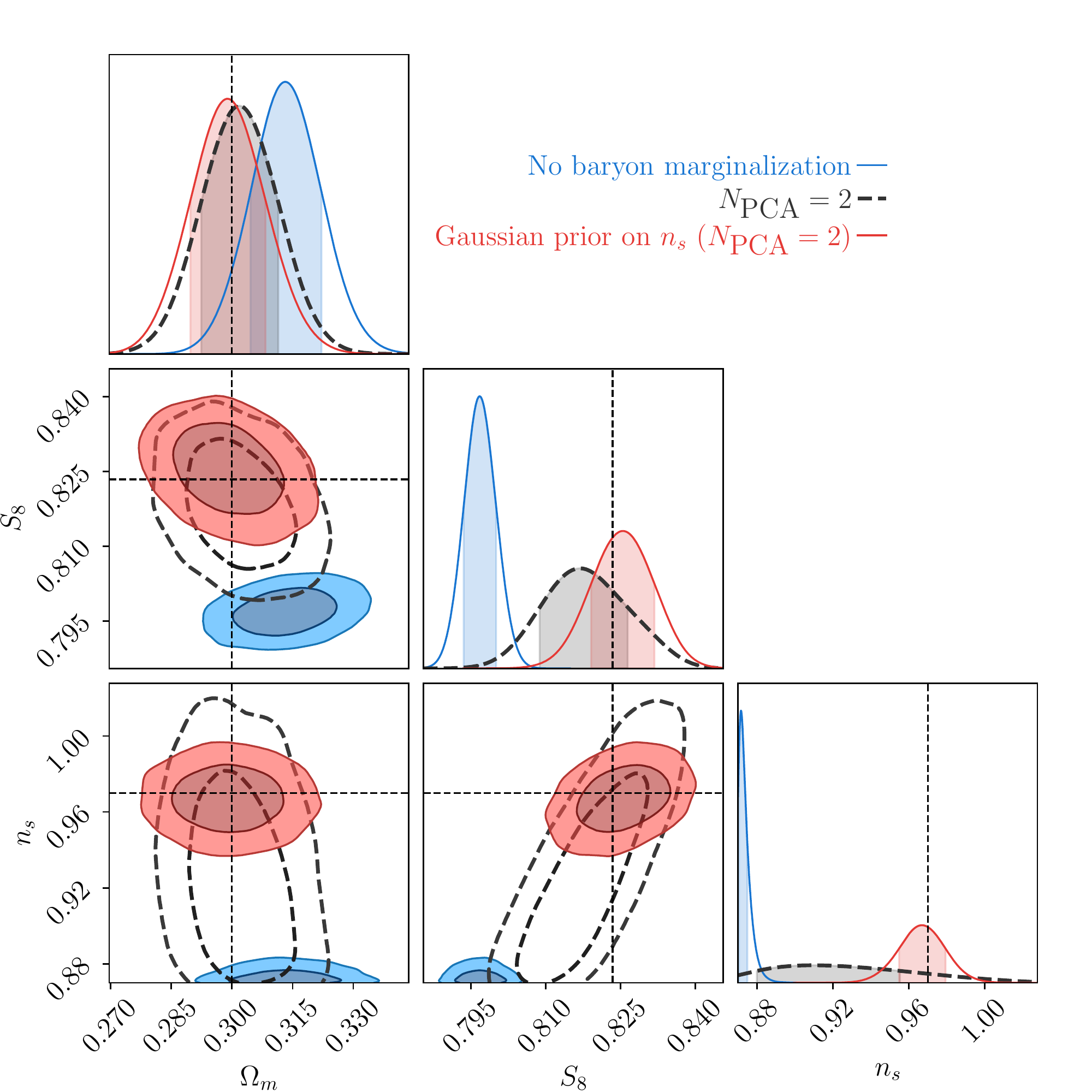}
    \caption{Results of cosmological inference with the data vectors contaminated with the BAHAMAS-T8.0 feedback scenario. The blue contours show the results with no baryon marginalization. The black dashed contours show the results where the baryonic effects are maginalized using the PCA marginalization scheme with 2 PCA modes. The red contours are the results with 2 PCA modes with an additional Gaussian prior on \ns. See section \ref{ssec:baryon_scale_cuts} for more details.
    }
    \label{fig:baryon_contours}
\end{figure}

\begin{figure*}
    \centering
    \includegraphics[width=\linewidth]{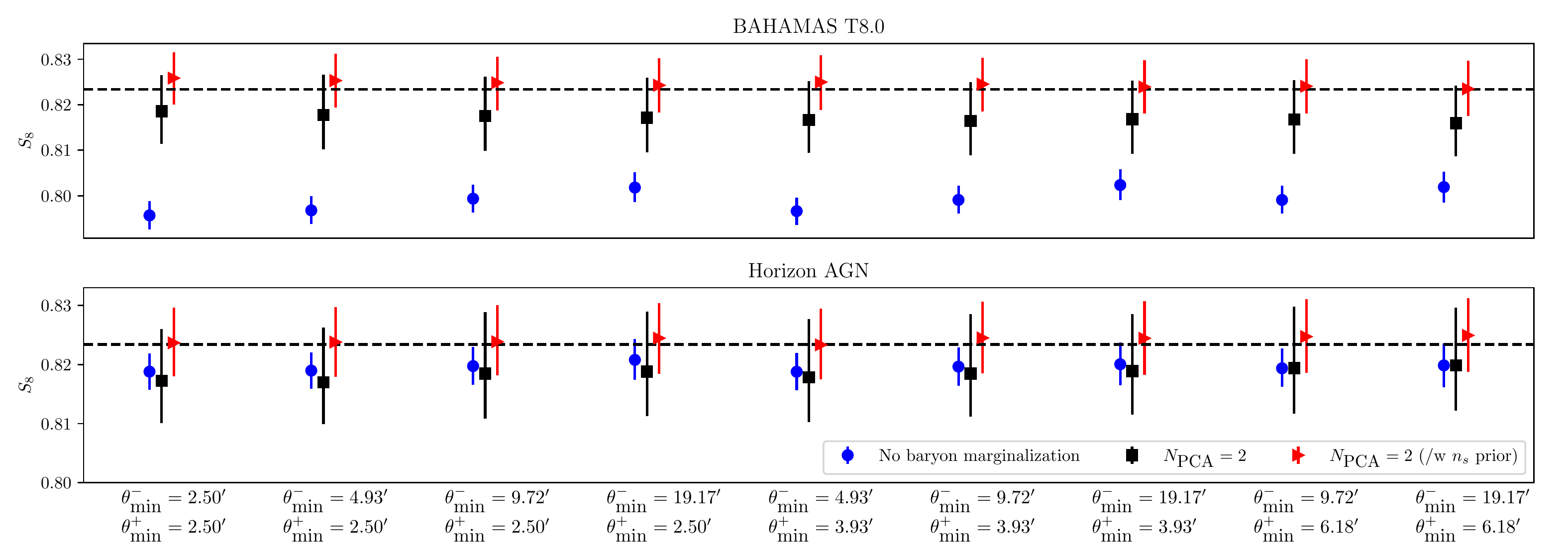}
    \caption{Marginalized one-dimensional $S_8$ posteriors for BAHAMAS T8.0 (top panel) and the Horizon AGN (bottom panel) feedback scenarios with different scale cuts. We show the results without any marginalization of baryonic effects (blue circles), using PCA marginalization with 2 PC modes (black squares) and using PCA marginalization with 2 PC modes with an added Gaussian prior on \ns (red triangles). The inferred value of $S_8$ is biased low without any mitigation for the baryonic effects. This is especially apparent for the BAHAMAS-T8.0 scenario. PCA marginalization is largely successful in recovering the value of $S_8$ in an unbiased manner, with a residual $\lesssim 1\sigma$ bias. The residual biases can be mitigated by adding an external prior on \ns. }
    \label{fig:parameter_bias}
\end{figure*}

In section \ref{ssec:gp_vs_nn} we show that for our NN emulator the majority of the time is spent on the calculation of the data vector and training the neural network. After training the NN emulator, any subsequent likelihood analysis can be done extremely quickly. For example, the LSST-Y1 3x2 point analysis ran in $\sim$ 5 minutes. In this section, we demonstrate the usefulness of such a pre-trained NN using the example of deriving scale cuts in the data vector due to baryonic physics effects in 3x2 point analyses. 

To study the impact of baryonic feedback on 3x2 point analyses, we create contaminated data vectors using the power spectra of hydrodynamic simulations. We can then treat each of the contaminated data vectors as the observed data vector in our analyses and investigate if our mitigation scheme leads to unbiased cosmological constraints. 

\citet{Huang2021} constrained different baryonic physics scenarios with DES-Y1 data and found that BAHAMAS-T8.0 is one of the strongest baryonic feedback scenario that is still compatible with observations. In addition to the BAHAMAS-T8.0 simulations we also create a synthetic data vector based on the Horizon-AGN physics scenario. We use these two feedback scenarios to study the impact of baryonic effects on our LSST-Y1 3x2 pt analysis. 

Recall that the emulator is trained to predict all the data vector elements without any scale cut. This allows us to impose different scale cuts on the emulator after it is trained and investigate the impact of different scale cuts.  

In Fig. \ref{fig:baryon_contours}, we show the cosmological contours for $\Omega_m$, $S_8$ and $n_s$ using the BAHAMAS-T8.0 contaminated data vector. We see that without any baryon mitigation scheme (blue contours), the inferred parameters can be highly biased, especially in the parameters $S_8$ and \ns. The black dashed contours show the results when using 2 PCA modes to mitigate the baryonic effects. 

While the PCA mitigation scheme can largely mitigate the biases in the inferred cosmological parameters, there is still a residual $\sim 1\sigma$ bias in the inferred value of $n_s$. Since the value of \ns is well constrained by cosmic microwave background (CMB) experiments, we test a Gaussian prior on \ns centred on the fiducial value, \ns$=0.97$ and a standard deviation, $\sigma[n_{\text{s}}]=0.012$, which is 3 times the width of the posterior on \ns inferred from the Planck data \citep{PlanckCosmoPars} (see red contours in Fig. \ref{fig:baryon_contours}). There is no residual bias in any parameters when using this modest prior.

In Fig. \ref{fig:parameter_bias}, we plot the one-dimensional marginalized $S_8$ error bars for the two different data vectors contaminated with BAHAMAS-T8.0 and Horizon-AGN. We consider 9 different scale cuts and show the results with no baryon mitigation (blue circles) and marginalization using 2 PC modes with (red triangles) and without (black square) the Gaussian prior on \ns. 

For the Horizon-AGN baryonic feedback scenario, depending on the scale cut, the inferred value of $S_8$ is biased low by $1$-$1.5\sigma$ if no mitigation scheme is applied. For the more extreme scenario of BAHAMAS-T8.0, without any mitigation of the baryonic feedback, the inferred value of $S_8$ is biased low by up to $9\sigma$. The biases in these parameters can be largely mitigated using the PCA marginalization scheme, where again we find that the modest prior on \ns provides a significant boost to the precision of the inferred $S_8$ value. 

We conclude that for LSST-Y1 3x2 point analyses, PCA marginalization along with an external prior on \ns is a promising strategy to mitigate baryonic physics effects even when including scales down to $\theta^{\pm}_{\text{min}} = 2.5^{\prime}$. Most importantly we note that altogether the 54 likelihood analyses contained in Fig. \ref{fig:parameter_bias} only take 5.5 hours runtime if they are run sequentially. Parallelizing all runs, which is trivial, would allow the analyst to determine scale cuts after only $\sim$5 mins. 

\section{Conclusion}\label{sec:conclusion}

We are entering the era of Stage-IV cosmological surveys that have the potential to uncover some of the long standing mysteries in cosmology. The analysis of these data sets requires accurate modeling of systematic effects and precise characterization of tensions between different probes and experiments.

To get the most out of these experiments, their survey design and analysis choices need to be optimized for the different science cases. Such optimization studies require significant computational resources due to the large number of expensive simulated analyses that are needed to understand the impact of various analysis choices. This aspect will become even more important for more complex physics models, beyond LCDM.

In this paper we present a strategy to massively accelerate cosmological likelihood analyses by emulating the theoretically computed model vectors. Given that building accurate emulators in high dimensional parameter spaces is challenging, we design an iterative scheme that selects new training points from a tempered posterior of the previous run. These training points (cosmological parameters) are used to compute new model vectors  yielding a larger sample of training data. 

We consider an LSST Y1 3x2 point analysis as an example and separate the cosmological and systematics parameter space into fast and slow parameters. The former category refers to parameters that act additively and multiplicatively on the data vector entries, the latter denotes parameters that enter through complex calculations. We emulate all calculations that are slow and a posteriori compute the effects of the fast parameters.
Our analysis uses a $\Lambda$CDM cosmological model along with modelling of systematics such as intrinsic alignment, photo-$z$, multiplicative shear calibration bias, galaxy bias and baryonic effects. 

Using our iterative scheme of adding new training data in the parameter regions of high posterior probability, we find that our emulator's results are hardly distinguishable from a brute force likelihood analysis after only 3 iterations. Additional active learning schemes such as Bayesian optimization \citep{Leclercq2018, Rogers2019} can be used to further improve the acquisition of new training points. 

We study two different types of emulators in this paper, one based on Gaussian process regression and a second one based on a simple neural net. Each of these emulators computes accurate cosmological results while speeding up the analysis by an order of magnitude compared to a highly optimized code running a brute force MCMC analysis. 

More importantly however, once the emulator has been trained, it can be reused with a runtime of only a few minutes. This allows future experiments to run the simulated likelihood analyses required to optimize survey design, determine analysis choices for the different science cases, and to accurately model tensions between probes and experiments. The number of simulated analyses required can easily range into several thousands for just one release of LSST data. 

We demonstrate this capability of our emulator by studying the impact of baryonic feedback on LSST-Y1 3x2 point analysis. For such an analysis, after running the full iterative training with the emulator once, we re-use the trained emulator to assess the impact of baryonic feedback for different scale cuts, mitigation approach and priors leading to $\mO(1000)$ times speed-up of the inference, in our case it takes approximately 5 minutes per simulated analysis.

Emulators must be re-trained when considering a different parameter space and when considering different cosmological probes. For a given parameter space and probe combination however any simulated analysis can be run in minutes. This solves a long standing bottleneck in cosmological analyses; experiments can now efficiently map out the science return as a function of analysis choices and survey strategy and at the same time reduce the carbon footprint of computationally expensive likelihood analyses.


\section*{Acknowledgement}
The computation presented here was performed on the High Performance Computing (HPC) resources supported by the University of Arizona TRIF, UITS, and Research, Innovation, and Impact (RII) and maintained by the UArizona Research Technologies department.
This paper is supported by the Department of Energy Cosmic Frontier program, grant DE-SC0020215.
\section*{Data availability statement}
The data products generated in this work will be shared on reasonable request to the authors.


\bibliographystyle{mnras}
\bibliography{emulator} 


\appendix
\section{Details of emulator design}\label{app:emu_details}

In this section, we present some more details of our emulator design.
\subsection{Input Normalization}\label{ssec:input_norm}

Training the emulator is easier if all the elements of the predicted quantity are $\mO(1)$. Therefore, we normalize the data vectors as,
\begin{equation}\label{eqn:dv_normalization}
    \mvec{\mD} = \frac{\mvec{d} - \mvec{d}_{\text{fid}}}{\mvec{\sigma}_d},
\end{equation}
where, $\mvec{d}_{\text{fid}}$ is the model data vector at some fiducial parameter values. $\mvec{\sigma}_d$ is the theoretically predicted standard deviation of the data vectors calculated from the covariance matrix, $\mmat{C}$,
\begin{equation}\label{eqn:sigma_d}
    \sigma_d = \sqrt{\text{diag}[\mmat{C}]}.
\end{equation}
We train our emulators to predict the normalized data vector $\mvec{\mD}$ instead of $\mvec{d}$. Furthermore, we normalize the input parameters as,
\begin{equation}
    \tilde{\mvec{\theta}} = \frac{\mTheta - \mTheta_{\text{min}}}{\mTheta_{\text{max}} - \mTheta_{\text{min}}},
\end{equation}
where, $\mTheta_{\text{min/max}}$ are the minimum and maximum value of the parameter in the training sample. Therefore, all the elements of $\tilde{\mvec{\theta}}$ are $\mO(1)$.

\subsection{Fast parameters}\label{ssec:fast_parameters}
In cosmological applications, the data vector depends on some parameters in a simple multiplicative or additive manner. For example, linear galaxy bias and multiplicative shear calibration parameters are simple multiplicative terms for calculating the data vectors. We call such parameters collectively as the `fast parameters'. Let us denote the multiplicative parameters as $\mTheta_{\text{mult}}$ and the additive parameters as $\mTheta_{\text{add}}$. 

The dependence on other parameters can be highly non-linear. We denote such parameters as $\mTheta_{\text{NL}}$. In our emulator design, we use our emulator to only model the dependence of the data vector on the non-linear parameters, $\mTheta_{\text{NL}}$. During the emulation, $\mTheta_{\text{mult/add}}$ are set to some fiducial value, $\mTheta^{\text{fid}}_{\text{mult/add}}$. The effect of the multiplicative and the additive parameters are calculated perfectly during the MCMC sampling as 
\begin{align}
    \mvec{d}_{\text{model}}(\mTheta) &= f(\mTheta_{\text{mult}}) \times \mvec{d}_{\text{emu}}(\mTheta_{\text{NL}}; \mTheta_{\text{mult}}=\mTheta^{\text{fid}}_{\text{mult}}), \\
    \mvec{d}_{\text{model}}(\mTheta) &= \mvec{d}_{\text{emu}}(\mTheta_{\text{NL}}; \mTheta_{\text{add}}=\mTheta^{\text{fid}}_{\text{add}}) + g(\mTheta_{\text{add}}) ,
\end{align}
where, $f$ and $g$ are simple functions of the multiplicative and the additive parameters. 

Modelling the dependence of the fast parameters in this way captures the exact dependence of $\mTheta_{\text{mult/add}}$. Therefore, it reduces the dimensionality of the input parameters, thus simplifying the emulation. While we follow this simplification for our emulation, we have checked that emulating using the full parameter space gives qualitatively similar results. 
\subsection{Scale cuts in emulation}\label{ssec:scale_cuts}
The theoretical modelling of $\mvec{d}_{\text{model}}$ may be impacted by deficiency in the small-scale modelling, e.g, due baryonic feedback and non-linear galaxy bias. To mitigate these impacts, the small scales are masked out in the inference in equation \eqref{eqn:chi_sq}. In our emulator, we use the scale cuts while calculating the log-likelihood. But while emulating, we predict the data vectors without any scale cuts. This allows us to re-use the emulator to test the impact of different scale cuts as we show in section \ref{ssec:baryon_scale_cuts}.
\section{Gaussian Process regression}\label{app:gp}

In this section, we present the technical details of Gaussian process regression. Gaussian process regression is a non-parametric Bayesian method where it is assumed that the function values, $\mvec{f} = \{f_1, f_2, \dots, f_N\}$, at $N$ points, $\{\mvec{x}\} = \{\mvec{x}_1, \mvec{x}_2, \dots, \mvec{x}_N\}$, are Gaussian distributed such that,
\begin{equation}\label{eqn:log_prob_gp}
    \mP(\mvec{f}|\mmat{C}_f) = \frac{1}{\sqrt{(2\pi)^{N}\text{det}(\mmat{C}_f)}}\exp \bigg[-\frac{1}{2}\mvec{f}^{\text{T}} \mmat{C}_f^{-1}\mvec{f}\bigg].
\end{equation}
Their covariance is modelled using a kernel function, $K$, that depends on the distance of the two points, such that,
\begin{equation}
    \mmat{C}^{mn}_{f} = \text{Cov}[f(\mvec{x}_m), f(\mvec{x}_n)] = K(\mvec{x}_m, \mvec{x}_n).
\end{equation}
For our GP emulator, we use a squared-exponential kernel which is given as,
\begin{equation}\label{eqn:gp_kernel}
    K(\mvec{x}_m, \mvec{x}_n) = a \exp\bigg[-\frac{1}{2}\sum_{i=1}^{N_{\text{dim}}}\frac{(x^i_m - x^i_n)^2}{b^2_i} \bigg],
\end{equation}
where, $a$ and $\mvec{b}$ are the hyper-parameters of the Gaussian process. The covariance of the function values decreases with increasing distance. Intuitively, this means that the function values of nearby points will be similar. The performance of GP regression depends strongly on the values of the hyper-parameters. To find the optimal value of the hyper-parameters, we maximize the log-probability, equation \eqref{eqn:log_prob_gp}, for the training sample. 

Once we have the optimized values of the GP hyper-parameters, we can use the Gaussian process to predict the function value at a new point. Assuming that the function value, $f^{*}$, at the new point, $\mvec{x}^{*}$, is also Gaussian distributed along with the function values of the training set, $f^{*}$ can be estimated using a conditional Gaussian distribution,
\begin{equation}
    \mP(f^{*}|\mvec{f}, \{\mvec{x}\}) = \mathcal{N}(\mu, \sigma^2).
\end{equation}
In the above, $\mu$ gives the mean prediction and $\sigma$ gives an estimate of the uncertainty in the prediction. The GP predicted error is known to be unreliable for the emulation of deterministic functions. E.g, \citet{Mootoovaloo2020} found that propagating the uncertainty into inference with an GP-based power spectrum emulator performs worse than only using the mean estimate. Therefore, in our emulator, we only use the mean estimate to predict the data vectors and discard the uncertainty in the GP estimates.  
The mean and the standard deviation of the predicted function values is given as,
\begin{align}
    \mu &= \mvec{K}^{\text{T}}_{*} \mmat{C}_f^{-1} \mvec{f}, \\
    \sigma^2 &= k_{*} - \mvec{K}^{-1}_{*} \mmat{C}_f^{-1} \mvec{K}_{*},
\end{align}
where, $\mvec{K}_{*}$ is the covariance between the function value at the new point, $f^{*}$, and the function values for the training data, $\mvec{f}$, such that,
\begin{equation}
    \mvec{K}^{i}_{*} = K(\mvec{x}^i, \mvec{x}_{*}),
\end{equation}
and
\begin{equation}
    k_{*} = K(\mvec{x}_{*}, \mvec{x}_{*}).
\end{equation}
Training and prediction with a Gaussian Process has $\mO(N_D^3)$ complexity, where $N_D$ is the size of the training data set. Therefore, Gaussian process regression cannot scale well with the size of the training sample. 
\section{Full 2D posterior}\label{sec:full_2D_posterior}

\begin{figure*}
    \centering
    \includegraphics[width=\linewidth]{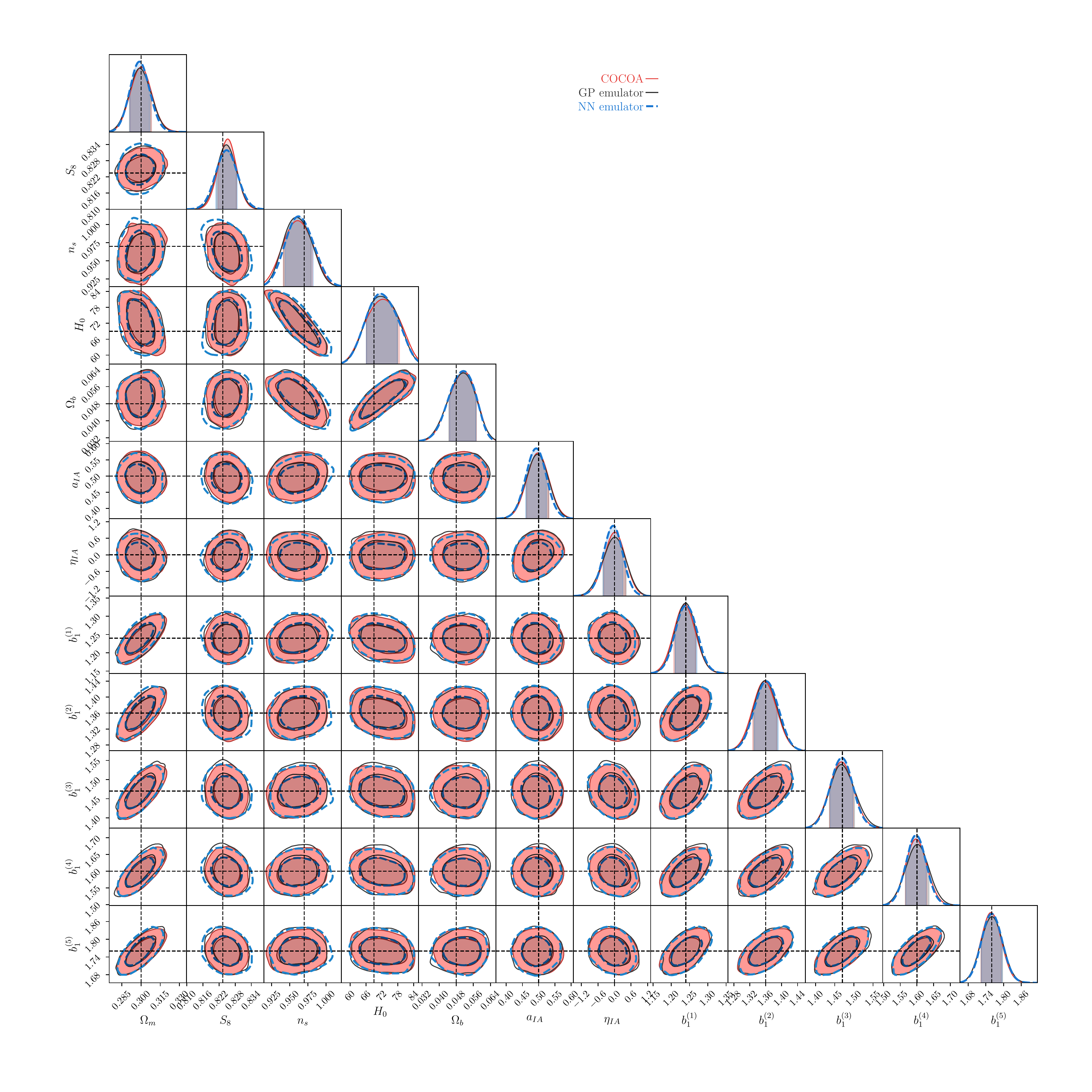}
    \caption{Triangle plot of 2D projected posteriors derived using \cocoa {\it (red contours)}, GP emulator {\it (black contours)} and the NN emulator {\it (blue dashed contours)}. We plot the contours for parameters that are likelihood-dominated, namely, the cosmological parameters, intrinsic alignments parameters and the linear galaxy bias parameters. The contours derived using the emulators are remarkably similar to the contours derived using the full theoretical model of \cocoa.}
    \label{fig:multidimensional_posterior}
\end{figure*}

In Figure \ref{fig:multidimensional_posterior}, we show the 2D projected posteriors inferred using our iterative emulator along with the posterior inferred using the full theoretical model with \cocoa. We plot the contours for the cosmological parameters, intrinsic alignment parameters and the galaxy bias parameter. The posteriors obtained using our emulators are in excellent agreement with the contours obtained from the full theoretical model of \cocoa.

\bsp	
\label{lastpage}
\end{document}